\documentclass[epjST]{svjour}
\usepackage{epsf}
\usepackage{latexsym}
\usepackage{amsmath}
\usepackage{nicefrac}
\usepackage{dsfont}

\newcommand{\be}{\begin{equation}}
\newcommand{\ee}{\end{equation}}
\newcommand{\bea}{\begin{eqnarray}}
\newcommand{\eea}{\end{eqnarray}}

\begin{document}
\title{Forward and inverse problems in fundamental and applied magnetohydrodynamics}
\author{Andre Giesecke\inst{1}\fnmsep\thanks{\email{a.giesecke@hzdr.de}} \and Frank Stefani \inst{1} 
\and Thomas Wondrak \inst{1}  \and Mingtian Xu \inst{2}}
\institute{Helmholtz-Zentrum Dresden-Rossendorf, P.O. Box 510119,
D-01314 Dresden,
Germany \and Shandong University, P. O. Box 88, 
Jing Shi Road 73 , Jinan City, Shandong Province, P. R. China}
\abstract{
This Minireview summarizes the recent efforts to solve forward and 
inverse problems as they occur in different branches of fundamental 
and applied magnetohydrodynamics. As for the forward problem, the main 
focus is on the numerical treatment of induction processes, 
including self-excitation of magnetic fields in non-spherical 
domains and/or under the influence of non-homogeneous material 
parameters. As an important application of the developed numerical 
schemes, the functioning of the von-K\'{a}rm\'{a}n-sodium (VKS) dynamo experiment 
is shown to depend crucially on the presence of soft-iron impellers. 
As for the inverse problem, the main focus is on the mathematical background 
and some first practical applications of the Contactless Inductive 
Flow Tomography (CIFT), in which flow induced magnetic field perturbations 
are utilized for the reconstruction of the velocity field. The 
promises of CIFT for flow field monitoring in the continuous 
casting of steel are substantiated by results obtained at a test rig with a 
low melting liquid metal. While CIFT is presently restricted to flows with 
low magnetic Reynolds 
numbers, some selected problems of non-linear inverse dynamo theory, 
with possible application to geo- and astrophysics, are also discussed.
} 
%
\maketitle
\section{Introduction}
\label{intro}
When a moving electrically conducting fluid comes under the influence of a 
magnetic field, an electromotive force is produced that drives electrical 
currents in the fluid which, in turn, modify the original magnetic field.
This general principle applies both to the case that a magnetic 
field is applied from outside, as well as to the case that
it is exclusively produced inside the fluid. In the latter case, we 
speak about self-excitation, or homogeneous dynamo action. It can only occur
if the magnetic Reynolds number $Rm=\mu \sigma L V$, i.e. the
product of magnetic permeability $\mu$, electrical conductivity $\sigma$,
size $L$ and velocity scale $V$ of the moving fluid, is
significantly greater than one. This condition is well fulfilled in 
the convecting fluid layers of cosmic bodies, such as the 
Earth \cite{WICHT} or the Sun \cite{JONES}, 
but it is much harder to achieve in the liquid metal laboratory. Yet, 
the threshold of self-excitation has been 
reached in some dedicated liquid sodium experiments in Riga \cite{PRL1}, 
Karlsruhe \cite{KARLSRUHE}, and Cadarache \cite{CADARACHE},
and it is targeted in a few more places around the world \cite{ZAMM}.
If, however, $Rm$ is not large enough for self-excitation, 
a magnetic field applied from outside will still be modified by
the moving fluid. 
This effect can be utilized for 
inferring the velocity field from the
externally measured induced magnetic field \cite{PRE}.

In this Minireview, we will deal with both effects: 
induction under the influence 
of applied magnetic fields, and self-excitation. 
We will discuss two 
mathematical formulations, the first one, the differential equation approach,
being based on the induction equation, the second one, 
the integral equation approach,
relying on Biot-Savart's law. 
In either case, particular attention will be paid to the correct 
treatment of the boundary 
conditions for the magnetic field, which is a notorious and 
non-trivial  problem in non-spherical geometries.
We will exemplify both methods by treating forward problems
related to the French von-K\'{a}rm\'{a}n-Sodium dynamo experiment for
which we will show that the use of magnetic propeller materials plays
a decisive role for the functioning and the mode selection of this
type of dynamo \cite{GIESECKENJP,GIESECKEPRL}.

The integral equation method is then applied to various  inverse 
induction problems in 2D and 3D. We will show that the Contactless
Inductive Flow Tomography (CIFT), as we call it now \cite{PRE}, has a 
promising potential
for the online flow monitoring in
technologies such as steel casting or crystal growth. 
The much harder, since non-linear problem of inverse dynamo theory 
is only touched upon by discussing some simplified models.

\section{Induction and self-excitation: Mathematical basics}
In this section we will discuss two different approaches to 
induction and self-excitation, the first on based on differential
equations, the second one on integral equations.
\subsection{Differential equation approach}
\subsubsection{The induction equation}
The evolution equation for the magnetic field ${\bf B}$ in a
fluid of electrical conductivity $\sigma$ and relative permeability $\mu_{r}$ 
can be derived from
Amp\`ere's law, Faraday's law, the divergence-free 
condition for the
magnetic field, and Ohm's law in moving conductors:
\begin{eqnarray}
\nabla \times \frac{{\bf{B}}}{\mu_0 \mu_{r}}= {\bf{j}} \label{1} \\
\nabla \times {\bf{E}}= -\frac{\partial {\bf{B}}}{\partial t} \label{2}\\
\nabla \cdot {\bf{B}}=0 \label{3}\\
{\bf{j}}=\sigma ({\bf{E}}+{\bf{v}} \times {\bf{B}}) \label{4}.
\end{eqnarray}
In Eq. (2) we have skipped the 
displacement current 
because  in good electrical conductors                              
the quasi-stationary approximation  can be applied. 
We will also restrict ourselves to
the case without external currents.
That way, taking the $curl$ of Eq. (1)  and Eq. (4), and
inserting
Eq. (2), we obtain 
the {\it induction equation}
for the magnetic field:

\begin{eqnarray}
\frac{\partial\vec{B}}{\partial t}
= \nabla\!\times\!\left(\!\vec{u}\!\times\!\vec{B}
+\frac{1}{\mu_{r}\mu_0\sigma}\frac{\nabla\mu_{r}}{\mu_{r}}\!\times\!\vec{B}
-\frac{1}{\mu_{r}\mu_0\sigma}\!\nabla\!\times\!\vec{B}\!\right)\!.\label{eq::indeq}
\end{eqnarray}

This is the general form in which spatially varying 
conductivities and relative permeabilities are taken into account \cite{GIESECKENJP},
as it will indeed be considered in the later treatment of the
VKS dynamo experiment.
If the material parameters are constant in a given volume, 
the induction equation
reduces to the better known simple form:
\begin{eqnarray}
\frac{\partial {{\bf{B}}}}{\partial t}=\nabla
\times ({\bf{v}} \times {\bf{B}})
+\frac{1}{\mu_0 \mu_{r} \sigma} \Delta {\bf{B}} \label{6} \; .
\end{eqnarray}
The right hand side of Eq. (\ref{6}) describes the competition between
the diffusion and the advection of the magnetic field.  Comparing the
diffusion time-scale with the kinematic time-scale gives the most
important dimensionless number of magnetohydrodynamics, the magnetic
Reynolds number $Rm:=\mu_0 \mu_r \sigma L V$

Depending on the actual flow pattern, the values of the critical $Rm$,
at which the field starts to grow, are usually in the range of
10$^1$...10$^3$ \cite{ZAMM}.  Most flows in cosmic bodies in which $Rm$ is large
enough will act as dynamos, although there are a number of anti-dynamo
theorems excluding too simple structures of the velocity field or the
self-excited magnetic field \cite{COWLING}.

Throughout this paper, the velocity field will be considered as
given or, in the inverse problems, as an unknown which we will try
to reconstruct. That means, in turn, that no attempts will be made to
solve the Navier-Stokes equation in the MHD regime.

\subsubsection{Simulations in spheres}

The easiest geometry for treating the induction equation, together
with the demand that ${\bf B}$ is curl free in the exterior and
vanishes at infinity, is the spherical geometry.  Fortunately, planets
and stars are in first approximation spherical so that dynamo problems
for those bodies are comparably easy to solve.

Considering the solenoidal character of magnetic fields, 
$\nabla\cdot\vec{B}=0$,
and  the incompressibility 
of liquid metals, $\nabla\cdot\vec{v}=0$, 
both the flow and the magnetic field can  be represented in a 
toroidal-poloidal decomposition:
\begin{eqnarray}
\vec{v}(\vec{r})&=&\nabla\times\nabla\times s(\vec{r})\hat{\vec{r}}+\nabla\times
t(\vec{r})\hat{\vec{r}}, \label{8}\\
\vec{B}(\vec{r})&=&\nabla\times\nabla\times S(\vec{r})\hat{\vec{r}}+\nabla
\times T(\vec{r})\hat{\vec{r}}.\label{9}
\end{eqnarray}
Following the Bullard-Gellman formalism \cite{BULLARDGELLMAN}
the scalar functions $s,S,t,T$ can then be expanded in
terms of the spherical harmonics $Y_{m,l}(\vartheta, \varphi)$:
\begin{eqnarray}
S(r,\vartheta, \varphi)=\sum\limits_{l,m}S_{l,m}(r) Y_{l,m}(\vartheta, \varphi)
\mbox{ and }
T(r,\vartheta, \varphi)=\sum\limits_{l,m}T_{l,m}(r) Y_{l,m}(\vartheta, \varphi).
\label{10}
\end{eqnarray}
where  we denote the  degree and order of the spherical harmonics by
$l$ and $m$, respectively  

By inserting Eqs.~(\ref{8},\ref{9}) in Eq.~(\ref{6}), multiplying with
the conjugate complex $Y^{*}_{l,m}(\vartheta, \varphi)$, integrating
over $d\vartheta d\varphi$, and applying the orthogonality relation of
spherical harmonics, one can derive a coupled system of equations for
the defining scalars $S_{l,m}(r)$ and $T_{l,m}(r)$ which are second
order in $r$. More details of this procedure can be found, e.g., in
\cite{TCFD}.  What is interesting in our present context is that the
insulating boundary conditions, which require $\nabla \times
\vec{B}=0$ in the exterior, lead to two boundary conditions for the
scalar fields $S_{l,m}$ and $T_{l,m}$ at the outer radius $R$ that are
{\it diagonal} in $l$ and $m$:
\begin{eqnarray}
\frac{\partial S_{l,m}(r=R)}{\partial
  r}+lS_{l,m}(r=R)&=&0\nonumber\\
\\[-0.5cm]
T_{l,m}(r=R)&=&0.\nonumber
\end{eqnarray}
It is this diagonality that makes dynamo simulations in spherical geometry 
quite simple, and 
the spherical harmonics decomposition was indeed the method of choice
for many simulation of cosmic dynamos, but also
for the optimization of spherical dynamo experiments such as the 
MDX experiment in Madison \cite{FOREST}.

\subsubsection{Treating the boundary conditions for non-spherical domains}

The convenient treatment of the boundary conditions, which are
diagonal in the degrees and orders of the spherical harmonics,
does not extend from spherical to other geometries. For that reason 
the simulation of liquid sodium dynamo experiments, most of 
them working in cylindrical geometry, needs significantly higher 
numerical effort. 

For rather compact machines like the Karlsruhe dynamo experiment, with
its 52 spin generators and a cylindrical aspect ratio close to one, it
appears reasonable to utilize the convenience of spherical harmonics
expansion by embedding the actual cylindrical vessel into an enclosing
virtual sphere and to assume the space between this enclosing sphere
and the real cylinder to be filled with matter of low conductivity.
Choosing conductivity factors of 1, 100, and 1000 between the cylinder
and the embedding, a convergence towards a realistic critical $Rm$ was
obtained in \cite{RAEDLER1}.

While this embedding method works reasonably well for cylinders with
an aspect ratio close to one, its application to very long dynamos is
inconvenient due to the rather slow convergence.  Hence, for the Riga
dynamo another method has been employed \cite{KLUWER,PLASMA,ANRIGA}
which is based on the fact that taking the limit $\mu \sigma
\rightarrow 0$ the induction equation reduces to a Laplace equation.
While in the dynamo region the induction equation is solved, the
Laplace equation is solved in the exterior by a pseudo-relaxation
method.  The solutions in both domains are than matched by appropriate
interface conditions for the magnetic field.
The scheme has been used extensively for 
the prediction, optimization and
data analysis of the Riga dynamo experiment and
the resulting growth rates and frequencies were 
in remarkable agreement with the experimental 
ones. 

In comparison, the utilization of
simplified vertical-field conditions only (i.e., vanishing tangential
fields at the boundary, $\mathbf{B}^{\rm{tan}}=0$), turned out to
underestimate the critical $Rm$ of the dynamo by some 
20 per cent, an error margin that is not acceptable for 
predictions of expensive liquid sodium experiments.

Another method to treat insulating boundary conditions has been developed in
\cite{GUERMOND1,GUERMOND2,GUERMOND3}. The scheme uses a spectral
finite element method and was applied successfully to several problems
in connection with the VKS experiment
\cite{GIESECKENJP,GIESECKE2010A,GIESECKE2010B}
and to precession-driven dynamos in cylindrical domains \cite{NORE}. 
The starting point 
is the fact that the divergence-free and curl-free 
magnetic field in the non-conducting 
exterior can be expressed by a gradient of a scalar
magnetic potential, according to ${\bf B}_{ext}= -\nabla \Phi$. 
Then, the Finite Element approximation  for the 
magnetic field ${\bf H}={\bf B}/\mu_{r}$ in the conducting interior is matched to the solution of the
outer Laplace equation by using an Interior Penalty Galerkin method.
Typically, the outer spherical insulating 
domain is made ten times larger than the inner domain so that
the concrete form of the boundary conditions at the very outer
boundary becomes irrelevant.

A possibility to avoid the solution of the Laplace equation in 
the exterior is to utilize the mathematically equivalent 
boundary element method (BEM).
This procedure has been proposed in two papers by Iskakov et al. 
\cite{ISKAKOV1,ISKAKOV2}, and was recently modified and applied to
various dynamo problems \cite{GIESECKE2008,GISSINGER}.
We will present this method in some more detail, since
it seems to be most efficient for the simulation of dynamo
problems in arbitrary domains.

The method starts with the normal component of the
magnetic field at the boundary as
obtained from a finite volume method
in the interior.
The unknown tangential components of the magnetic field at
timestep $(n+1)$ can then be derived by a matrix operation on a 
vector composed of the normal components of ${\bf{B}}$ at all
surface elements of the computational domain. 

The base of the BEM is once again the possibility to express the
magnetic field $\vec{B}$ as the gradient of a scalar magnetic
potential $\Phi$ which fulfills the Laplace equation:
\be
\vec{B}=-\nabla\Phi \quad\mbox{  with  }\quad \Delta\Phi =0, \quad
\Phi \rightarrow O(r^{-2}) \mbox{ for } r\rightarrow\infty.
\label{eq::laplace}
\ee
The computation of $\vec{B}$ on the boundary 
therefore requires the
integration of $\Delta\Phi=0$.
Consider a volume $\Omega$ that is bounded by the surface 
$\Gamma$ where $\nicefrac{\partial}{\partial n}$ denotes the derivative in the
normal direction: $\nicefrac{\partial}{\partial n}=\vec{n}\cdot\nabla$ so that
$\partial_n\Phi=-B^{\rm{n}}$ yields the normal component of
$\vec{B}$ on $\Gamma$. 
We can write Greens second identity for the scalar 
magnetic potential
$\Phi$ and a test-function $G$: 
\be
\int\limits_{\Omega}\left(G\Delta \Phi-\Phi\Delta G \right)d\Omega
=\int\limits_{\Gamma}\left(G\frac{\partial \Phi}{\partial
  n}-\Phi\frac{\partial G}{\partial n}\right)d\Gamma.\label{eq::greens2nd}
\ee
Using $\Delta\Phi=0$,  the potential 
is determined by the integral expression 
\be
\Phi(\vec{r})=\int\limits_{\Gamma} \left(G(\vec{r},\vec{r}')\frac{\partial \Phi(\vec{r}')}{\partial
  n}-\Phi(\vec{r}')\frac{\partial G(\vec{r},\vec{r}')}{\partial n}\right)d\Gamma(\vec{r}')\label{eq::inteq}
\ee
where $G(\vec{r}, \vec{r}')$ is the Green's function, or fundamental solution,
given by
\be
G(\vec{r},\vec{r}')=\displaystyle-\frac{1}{4\pi\left|\vec{r}-\vec{r}'\right|}
\ee
and therefore fulfilling $\Delta G(\vec{r},\vec{r}')=-\delta(\vec{r}-\vec{r}')$.
In principle, Eq.~(\ref{eq::inteq}) is only valid for
$\vec{r}\notin \Gamma$. For all source points on the boundary
the integration domain is
enlarged by a small half sphere with the radius $\epsilon$
and a short calculation of the limit 
$\epsilon\rightarrow 0$ results in the boundary integral
equation (see e.g. \cite{GIESECKE2008}): 
\be
\frac{1}{2}\Phi(\vec{r})=\int\limits_{\Gamma}\bigg(G(\vec{r},
\vec{r}')\underbrace{\frac{\partial \Phi(\vec{r}')}{\partial
  n}}_{\displaystyle -B^{\rm{n}}(\vec{r}')}-\Phi(\vec{r}')\frac{\partial
  G(\vec{r},\vec{r}')}{\partial n}\bigg)d\Gamma(\vec{r}'). \label{eq::bie_phi}
\ee
From~(\ref{eq::bie_phi})
the tangential components of the magnetic field on the boundary
$B^{\rm{t}}=\vec{e}_{\tau}\cdot\vec{B}=-\vec{e}_{\tau}\cdot\nabla\Phi(\vec{r})$
are computed by:
\be
{B}^{\tau}=2\int\limits_{\Gamma}\vec{e}_{\tau}\cdot
\left(\Phi(\vec{r}')\nabla_{r}\frac{\partial
  G(\vec{r},\vec{r}')}{\partial
  n}+B^{\rm{n}}(\vec{r}')\nabla_{r}G(\vec{r},\vec{r}')\right)d\Gamma(\vec{r}'),
\label{eq::bie_b}
\ee
where $\vec{e}_{\tau}$ represents the tangential unit vector on the surface
element $d\Gamma(\vec{r}')$.

A discretization of the equation system ~(\ref{eq::bie_phi})
and~(\ref{eq::bie_b}) yields an algebraic system of equations which
allows for the computation of the (unknown) tangential components of
the magnetic field.  The discretization of $\Gamma$ in boundary
elements is realized by the tessellation resulting from the finite
volume discretization on the domain surface.

\subsection{Integral equation approach}

An alternative way to deal with the non-local boundary conditions of
MHD is the integro-differential formulation developed by Meir and
Schmidt \cite{MEIRSCHMIDT1,MEIRSCHMIDT2,MEIRSCHMIDT3}.  Using the
current density as the primary electromagnetic variable, it is
possible to avoid artificial boundary conditions and fully account for
the electromagnetic interaction between flow region and
surrounding. The point is that the magnetic field has then to be
expressed in terms of the current by Biot-Savart's law.  This method
is presently being used in a realistic simulation of liquid metal
experiment on the so-called Tayler instability, a current-driven kink
type instability \cite{SEILMAYER}.

Interestingly, this method can be extended into an iterative scheme,
by which not only induction effects, but also self-excitation can be
treated.  Starting with an applied field ${\bf B}_0$, the authors of
\cite{BOURGOIN} search for solutions of the total magnetic field ${\bf
  B}={\bf B}_0+{\bf B}_{ind}={\bf B}_0+\sum_{k=1}^{\infty}{\bf B}_k$,
with ${\bf B}_k={\cal O} ({Rm^k})$. Actually, from a given ${\bf B}_k$
they compute the electromotive force ${\bf u} \times {\bf B}_k$, from
which the Poisson equation for the electric potential of the next
iteration $\Delta \varphi_{k+1}=\nabla \cdot ({\bf u} \times {\bf
  B}_k)$ is derived.  From this, the electric current of the $i+1$
iteration is obtained according to ${\bf j}_{k+1}=-\nabla
\varphi_{k+1}+{\bf u} \times {\bf B}_k$, and from this current the
magnetic field of the $i+1$ iteration is obtained via Biot-Savart's
law
\be
{\bf B}_{k+1}({\bf r})=\frac{Rm}{4 \pi} \int d^3 r'
\frac{{\bf j}_{k+1}({\bf r'})\times ({\bf r}-{\bf r'})}{|{\bf r}-{\bf r'}|^3}.
\label{biot}
\ee

Despite the convergence problems of this method when approaching the
critical $Rm$ of a dynamo, it is quite helpful to illustrate the
various induction processes that enter the dynamo mechanism.

With view on Eq. (\ref{biot}) it is tempting to ask why not to
consider Biot-Savart's law as an integral equation for the magnetic
field which has to be solved self-consistently.  Assuming an infinite
domain of homogeneous conductivity, such an approach has been utilized
early by Gailitis \cite{GAILITISINTEGRAL} for the dynamo action of a
pair of annular vortices and later by Dobler and R\"adler
\cite{DOBLERRAEDLER} for various dynamo problems.

An extension of such an integral equation method which incorporates
the effect of boundaries can already be found in the book of Roberts
\cite{ROBERTSBOOK}.  However, real use of the method was only made
later in the papers \cite{AN,JCP}.  With \cite{PREMINGTIAN} this
integral equation method was then extended to the case of unsteady
magnetic fields which requires an additional equation for the vector
potential $\vec{A}$ which is related to the magnetic field by
$\mathbf{B}=\nabla\times \mathbf{A}$. 
The general formulation of the integral equation
approach is the following:

\begin{eqnarray}
{\mathbf{b}}({\mathbf{r}})&=&\frac{\mu\sigma}{4\pi}\int_V\frac{({\mathbf{u}}({\mathbf{r}}')
\times({\mathbf{B}}_0({\mathbf{r}}')+{\mathbf{b}}({\mathbf{r}}')))\times({\mathbf{r}}-{\mathbf{r}}')}
{|{\mathbf{r}}-{\mathbf{r}}'|^3}dV'\nonumber \\ &&-\frac{\mu\sigma}{4\pi}\int_S\phi({\mathbf{s}}'){\mathbf{n}}({\mathbf{s}}')\times
\frac{{\mathbf{r}}-{\mathbf{s}}'}{|{\mathbf{r}}-{\mathbf{s}}'|^3}dS' 
-\frac{\mu\sigma\lambda}{4\pi}\int_V\frac{{\mathbf{A}}({\mathbf{r}}')
\times({\mathbf{r}}-{\mathbf{r}}')}{|{\mathbf{r}}-{\mathbf{r}}'|^3}
dV'\label{int1}\\
\frac{1}{2}\phi({\mathbf{s}})&=&\frac{1}{4\pi}\int_V\frac{({\mathbf{u}}({\mathbf{r}}')
\times({\mathbf{B}}_0({\mathbf{r}}')+{\mathbf{b}}({\mathbf{r}}')))\cdot({\mathbf{s}}-{\mathbf{r}}')}
{|{\mathbf{s}}-{\mathbf{r}}'|^3}dV'\nonumber \\  
&&-\frac{1}{4\pi}\int_S\phi({\mathbf{s}}'){\mathbf{n}}({\mathbf{s}}')\cdot\frac{{\mathbf{s}}-{\mathbf{s}}'}{|{\mathbf{s}}-
{\mathbf{s}}'|^3}dS'
-\frac{\lambda}{4\pi}\int_V\frac{{\mathbf{A}}({\mathbf{r}}')
\cdot({\mathbf{s}}-{\mathbf{r}}')}{|{\mathbf{s}}-{\mathbf{r}}'|^3}dV'\label{int2}\\
{\mathbf{A}}({\mathbf{r}})&=&\frac{1}{4\pi}\int_V\frac{({\mathbf{B}}_0({\mathbf{r}}')+{\mathbf{b}}({\mathbf{r}}'))
\times({\mathbf{r}}-{\mathbf{r}}')}{|{\mathbf{r}}-{\mathbf{r}}'|^3}dV'\nonumber \\
&&+\frac{1}{4\pi}\int_S{\mathbf{n}}({\mathbf{s}}')\times\frac{{\mathbf{B}}_0({\mathbf{s}}')+{\mathbf{b}}({\mathbf{s}}')
}{|{\mathbf{r}}-{\mathbf{s}}'|}dS',\label{int3}
\end{eqnarray}
where ${\mathbf{B}_0}$ is the externally applied magnetic field (which might be zero),
${\mathbf{b}}$  the induced magnetic field,
${\mathbf{u}}$ the
velocity field, ${\mathbf{A}}$ the vector potential, $\phi$ the
electric potential and ${\mathbf{n}}$ denotes the outward directed unit
vector at the boundary $S$. For a steady velocity field, the time
dependence of all electromagnetic fields can be assumed to be
$\sim \exp{\lambda t}$. 
In (\ref{int1}--\ref{int3}) the temporal variations of the velocity field
are ignored, which is justified when the typical diffusive timescale of
the magnetic field is shorter than the timescale of velocity
variations (which usually is the case in the context of liquid metal
experiments).

We have to distinguish three different cases: For non-zero
$\mathbf{B}_0$, and below the self-excitation threshold, the imaginary
part of $\lambda$ is simply the angular frequency of the applied and
also of the induced magnetic field. For $\mathbf{B}_0=0$ the equation
system (\ref{int1}--\ref{int3}) represents an eigenvalue equation for
the unknown time constant $\lambda$ whose real part is the growth
rate, and its imaginary part the angular frequency of the eigenmodes.
For $\mathbf{B}_0=0$ and $\lambda=0$, we need only Eqs. (\ref{int1})
and (\ref{int2}) which then reduce to an eigenvalue problem for the
critical value of the velocity $\mathbf u$ at which the dynamo starts
to work.

Note that the integral equation system (\ref{int1}--\ref{int3}) is by
far not the only possible one. There are indeed other possible
schemes, one of them starting from the Helmholtz equation for the
vector potential which leads, however, to a nonlinear eigenvalue
problem in $\lambda$ (see \cite{DOBLERRAEDLER}), while the above
scheme is a linear eigenvalue problem in $\lambda$ which has many
advantages when it comes to the numerical simulations.

The numerical implementation of the integral equation approach was applied to various dynamo
problems \cite{PREMINGTIAN,AMBI,RAUL,JCP2}, and helped also to
understand the under-performance of the first VKS experiment in terms
of the detrimental role of (rotating) lid layers behind the impellers
\cite{AMBI}.

\section{Forward problems: One example}

In this section we will invoke both the differential equation approach and
the integral equation approach for  understanding
the so-called von-K\'{a}rm\'{a}n-dynamo 
(VKS) experiment in Cadarache \cite{CADARACHE}, France.

The working principle of the VKS dynamo is illustrated in Fig 1a.
The flow of liquid sodium in a cylinder with diameter 412 mm and height
524 mm is produced by two counter-rotating impellers, each consisting
of a massive disk and 8 curved blades. 
The resulting flow is of the so-called s2$^+$t2 type, which comprises 
two poloidal eddies, indicated by ''s2'', with radial inflow  (+) in 
the equatorial plane, and two counter-rotating toroidal 
eddies, indicated by ''t2''. This
flow topology had been early shown to be a good candidate for dynamo
action \cite{DUDLEYJAMES}. 
In connection with the optimization of the VKS dynamo experiment,
Mari\'{e}, Normand and Daviaud \cite{MND}  had studied a rather 
simple analytical 
test flow of this topological type, which is expressed in the following way:
\begin{eqnarray}
u_r&=&-\frac{\pi}{2} \; r \; (1-r)^2 (1+2r) \cos (\pi z)\label{mnd1}\\
u_{\varphi}&=&4 \epsilon r (1-r) \sin (\pi z/2)\label{mnd2}\\
u_z&=&(1-r)(1+r-5 r^2) \sin (\pi z) . \label{mnd3}
\end{eqnarray}

In order to study dynamo action for this MND-flow, with correctly 
implementing the vacuum boundary conditions, we had utilized both the
differential equation approach (with solving the Laplace equation in 
the exterior)
as well the integral equation approach \cite{AMBI}. For 
the simplest case without any additional layers around the dynamo, 
and a toroidal/poloidal ratio $\epsilon=0.7259$, 
the structure of the resulting magnetic eigenfield is illustrated 
in Figs. 1b,c. In Fig. 1b the field lines of the equatorial dipole are seen,
while the isosurfaces  of 
the magnetic field energy (Fig. 1c) show the  
banana-like structure typical for this equatorial dipole eigenmode. 
\begin{figure}[h!]
\begin{center}
\epsfxsize=13cm{\epsffile{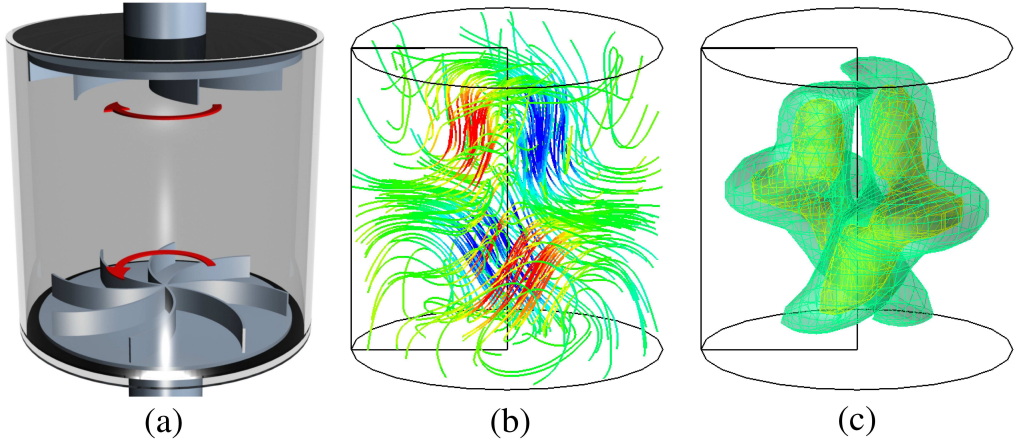}}
\end{center}
\caption{The VKS dynamo, and its numerical simulation by the integral 
equation method. 
(a) Sketch of the experimental setup. The von-K\'{a}rm\'{a}n swirling 
flow is driven by two counter-rotating
impellers.
(b) Magnetic field lines of the equatorial dipole field
resulting for the simplifying MND flow. (c) Iso-surface of the 
magnetic field energy.}
\end{figure}

A somewhat surprising results of \cite{AMBI} was 
that different kinds of layers around the flow field region 
can have opposite effects
on the critical magnetic Reynolds number. While a 
(stationary) side layer works usually in favor of dynamo action,
any layer (stationary or, even worse, rotating) behind the 
impellers impedes dynamo action significantly.
This observation was one of the motivations for the VKS group to 
replace the non-magnetic impeller material by soft-iron, 
connected with the idea that such a magnetic material  
might shield the detrimental induction effects 
from behind the impellers.

As a result of this modification, 
dynamo action was indeed observed \cite{CADARACHE}, and subsequent
experiments have brought about a remarkable variety of 
interesting phenomena, including field reversals, excursions, and
bursts \cite{CADARACHE2}. Such effects had been neither  
observed  in the Riga nor in the Karlsruhe experiment and thus instigated 
a strong interest in the geophysics community, with 
a number of attempts to understand reversals of the 
geomagnetic field and reversals in the VKS dynamo on the common 
basis  of low-dimensional models \cite{PETRELIS}.

Yet, the basic dynamo mechanism underlying such exciting
results remained unexplained for some time. In contrast to all
numerical simulations of the VKS dynamo, which had predicted a
non-axisymmetric (m=1) eigenmode similar to that shown in Fig. 1,
the actually observed eigenmode turned out to be more or 
less axisymmetric and dominated by the toroidal component. 
Furthermore, the experimentally observed critical $Rm$ around 30 was 
significantly lower than the predicted one 
of about 50. The question arose, therefore, whether both effects 
could be explained by the utilization of magnetic impellers.  

\begin{figure}[h!]
\begin{center}
\epsfxsize=13cm{\epsffile{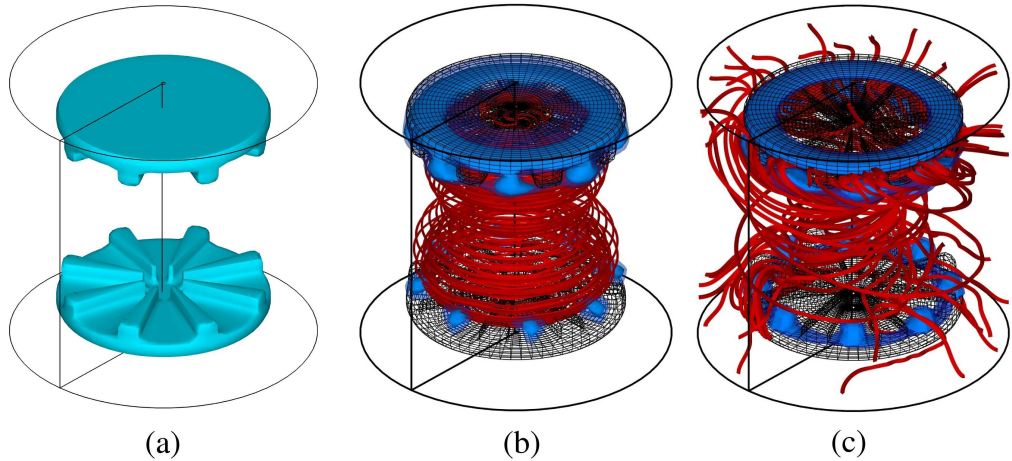}}
\end{center}
\caption{Effect of the magnetic impellers on the eigenmode selection in the VKS 
dynamo. (a) Assumed permeability
distribution; isolevel at 87.5 percent of the peak value. 
In the fluid region $\mu_r$ is equal to unity.
(b,c) Distribution of magnetic energy density and geometric
structure of the magnetic field for $\mu_r=20$.
(b) $Rm=30$, $\alpha=0$: decaying axisymmetric mode with vanishing
poloidal component. (c) $Rm=30$, $\alpha=-1.5$: Growing 
axisymmetric mode with poloidal component (after \cite{GIESECKEPRL}).}
\end{figure}

In order to answer this question, we have utilized the combined Finite
Volume/ Boundary Element Scheme exposed in 2.1.3.  With this code we
treated a slightly simplified model of the VKS dynamo, in which the
bended thin blades were replaced by straight thick ones (see Fig. 2a).
Our solutions show two different regimes that differ in the topology
of the leading eigenmode.  If the relative magnetic permeability of
the disks exceeds a threshold of around 20, we observe an interchange
of the leading eigenmode from the usual poloidal one to the toroidal
one \cite{GIESECKENJP}.  Yet, this magnetic disk, together with an
assumed MND flow in the bulk of the fluid, does not lead to dynamo
action, it just gives a toroidally dominated, but still decaying
eigenmode (Fig. 2b).  What is still needed for dynamo action is some
amount of helical turbulence, parameterized by the so-called $\alpha$
effect, which can well be assumed to exist in the space between the
blades \cite{RAVELET}. This $\alpha$ effect is then able to close the
dynamo loop by transforming toroidal field into poloidal field
components \cite{GIESECKEPRL}.  The resulting eigenmode, which can now
have a positive growth rate, is depicted in Fig. 2c.  The
experimentally observed eigenmode structure, with its strongly
dominant toroidal component, is much in favor of our numerical
simulation which explains the functioning of the VKS dynamo by means
of strong suction of magnetic flux into domains with enhanced
permeability (so called paramagnetic pumping \cite{DOBLER}), 
together with some $\alpha$ effect concentrated between
the blades.  This example shows that a very detailed treatment of
induction effects, with a careful consideration of gradients of
material parameters, might be essential for the understanding of
experimental results.

\section{Inverse problems: Some examples}
Assuming a velocity field as given, and asking for the 
resulting induced (or self-excited) magnetic field, 
represents a typical forward problem. 
However, the direction of interest can 
also be inverted: assume the magnetic field to be 
known from measurements in the exterior of the fluid, 
what is then the velocity that produces this field? 

\subsection{Inverse problems at low $Rm$}

Based on the integral equation system (\ref{int1}-\ref{int3})
in the time-independent limit (i.e. $\lambda \rightarrow 0$), 
we have developed  the so-called Contactless 
Inductive Flow Tomography (CIFT)
for the reconstruction of velocity fields from externally 
measured magnetic fields  \cite{PREMINGTIAN,IP1,IP2,MST}.
For the case of small $Rm$, which applies well to many
industrial problems like steel casting or silicon crystal growth,
the total field ${\bf B}={\bf B}_0+{\bf b}$ under the integrals in (\ref{int1}-\ref{int2})
can be replaced by the applied magnetic field  ${\bf B}_0$ alone.
This replacement makes the inverse problem of 
reconstructing the velocity field from the induced 
electro-magnetic field ${\bf b}$ a linear one, which is much 
easier solvable than the full-fledged non-linear 
problem with ${\bf B}={\bf B}_0+{\bf b}$ under the integrals.

In a first step \cite{IP1,IP2}, we had shown that, with a vertical 
field $B_{0,z}$  being applied, the velocity structure of 
the flow can 
be roughly reconstructed from the measurement of the 
induced magnetic field in the exterior and of the induced 
electric potential at the boundary of the fluid. There remains, 
however, a non-uniqueness concerning the exact radial 
distribution of the flow, a fact that can be made plausible 
by representing the fluid velocity by two defining scalars 
(for the poloidal and the toroidal velocity component), 
living both in the whole fluid volume. Then it is clear that 
two quantities measured only on a two-dimensional covering 
of the fluid cannot give enough information for the 
reconstruction of the two desired three-dimensional quantities.

Later, the reconstruction method was advanced to a completely contactless method,  
CIFT, to avoid the electric potential measurement at the 
fluid boundary \cite{MST}. The main idea was to apply the external 
magnetic field in two different 
directions (e.g. $B_{0,z}$ and $B_{0,x}$) and to utilize 
both corresponding sets 
of induced magnetic fields for the velocity field reconstruction.

\begin{figure}[h!]
\begin{center}
\epsfxsize=13cm{\epsffile{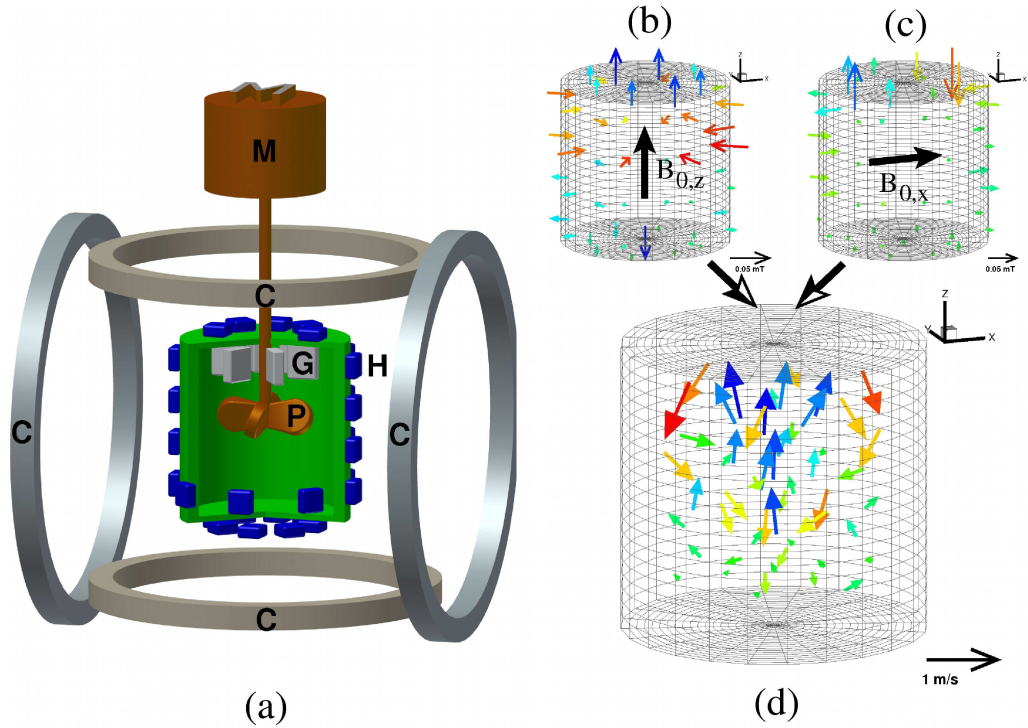}}
\end{center}
\caption{Application of CIFT to a 3D flow driven by a propeller
in a cylindrical vessel. (a) Schematic sketch of the
experiment: M - Motor, C - Coils, P - Propeller, G - Guiding blades, H - Hall sensors.
(b,c,d) Special case of upward pumping: (b) externally measured 
induced magnetic fields for applied $B_{0,z}$ and (c) $B_{0,x}$, (d) 
velocity field as reconstructed from the (b) and (c).}
\end{figure}

The goal of a first demonstration experiment \cite{PREMINGTIAN} for CIFT was the 
reconstruction of a propeller-driven three-dimensional flow of 
the eutectic alloy GaInSn in a compact cylindrical vessel with 
a ratio of height to diameter close to one (see Fig. 3a). In order 
to determine both the poloidal (in $r$ and $z$) and the toroidal (in $\varphi$)
flow components, we applied subsequently a vertical ($B_{0,z}$) and a 
horizontal ($B_{0,x}$) magnetic field, either produced by Helmholtz-like 
coil pairs. In this experiment, the switching between the two fields 
occurs every 3 seconds, so that after 6 seconds all the magnetic field information is available 
for the velocity reconstruction. Note that this 
rather poor time resolution can be significantly enhanced, with 
a physical limitation given by the magnetic decay time, 
which is in the order of 0.05 s for 
the demonstration experiment.
For each of the two applied fields, $B_{0,z}$ and $B_{0,x}$, the 
induced fields are measured by Hall sensors at 48 positions, 
which are rather homogeneously distributed around the surface 
of the cylindrical vessel. The small ratio of 
10$^{-3}$...10$^{-2}$ between the induced and the 
applied fields demands for a very stable current 
source of the Helmholtz-like coils, for a very stable 
relative position of Hall sensors and coils, and for 
significant effort to compensate drift and sensitivity changes of
the sensors.

For the reconstruction of the velocity field we solve the Gauss 
normal equations that minimize the mean quadratic deviation of 
the measured from the modeled induced magnetic fields \cite{PREMINGTIAN}. The 
intrinsic non-uniqueness problem concerning the detailed 
depth-dependence of the velocity is overcome by utilizing 
the so-called Tikhonov regularization by minimizing,
in parallel to the magnetic field deviations, also the kinetic 
energy of the flow or other appropriate quadratic functionals 
of the velocity. 
Mathematically, the Tikhonov regularization  
minimizes the total functional
\begin{eqnarray}
F[{\bf{v}}]=F_{B_{0,x}}[{\bf{v}}]+F_{B_{0,z}}[{\bf{v}}]+
F_{div}[{\bf{v}}]+F_{reg}[{\bf{v}}]
\label{tichfunc}
\end{eqnarray}
with
\begin{eqnarray}
F_{B_{0,x}}[{\bf{v}}]&=&\sum_{i=1}^{N_B}\frac{1}{\sigma^2_{i}}
\left( b_{i,meas}^{(B_{0,x})}-b_{i}^{(B_{0,x})}[{\bf{v}}]\right)^2
\label{eq8}\\
F_{B_{0,z}}[{\bf{v}}]&=&\sum_{i=1}^{N_B}\frac{1}{\sigma^2_{i}}
\left( b_{i,meas}^{(B_{0,z})}-b_{i}^{(B_{0,z})}[{\bf{v}}]\right)^2
\label{eq9}\\
F_{div}[{\bf{v}}]&=&\frac{1}{\sigma^2_{div}}
\sum_{k=1}^{N_V}\left(\nabla \cdot {\bf{v}}\right)^2_k \Delta V_k
\label{eq10}\\
F_{reg}[{\bf{v}}]&=&\frac{1}{\sigma^2_{pen}}
\sum_{k=1}^{N_V}{\bf{v}}^2_k \Delta V_k \; .
\label{eq11}
\end{eqnarray}
The two functionals in (\ref{eq8}) and in (\ref{eq9}) represent, for applied transverse field ${\bf B}_{0,x}$
and axial field  ${\bf B}_{0,z}$, respectively, the mean 
squared residual deviation of the measured induced 
magnetic fields $b_{i,meas}^{(B_{0})}$ from the modeled 
fields $b_{i}^{(B_{0})}[{\bf v}]$ (that result from the solution of
the corresponding forward problem). 
$F_{div}[{\bf{v}}]$ (Eq. \ref{eq10})
enforces the velocity field to be solenoidal,
and $F_{reg}[{\bf{v}}]$ (Eq. \ref{eq11}) is the regularization functional which 
tries to minimize the kinetic energy. 
The parameters $\sigma_{i}$ in (\ref{eq8}) and (\ref{eq9}) are the assumed 
a-priori errors for the measurement of the induced fields. 
The parameter $\sigma_{div}$ in (\ref{eq10}) is chosen very small as it
is a measure for the
divergence the velocity solution is allowed to have.  
The parameter $\sigma_{pen}$  in (\ref{eq11}) determines the trade-off between minimizing the
mean squared residual deviation of the observed fields and minimizing the
kinetic energy of the estimated velocity field. 
The normal equations, that follow from the minimization of the functional 
(\ref{tichfunc}), 
are solved by Cholesky decomposition.
This numerical scheme can easily be extended 
to incorporate further a-priori information, e.g., the velocity 
components or mass flow rates known from other measuring techniques.

In the model experiment it was possible to distinguish clearly 
between upward and downward pumping of the propeller, with the 
rotational component being reduced by guiding blades in the case 
of upward pumping. For this case, Figs. 3b,c show the induced 
fields for applied  $B_{0,z}$ and $B_{0,x}$, respectively, and the 
velocity field as reconstructed from this information (Fig. 3d). 
The comparison with 
UDV measurements has shown a good coincidence of the 
resulting velocity fields (see \cite{PREMINGTIAN}).

While CIFT is able to infer full 3D 
velocity fields by applying subsequently
two different external magnetic fields, for the following case related
to thin slab casting it can be reduced to the determination of the
velocity components parallel to the wide faces of the mould 
\cite{WONDRAK1,WONDRAK2,WONDRAK3}.
Figure 4a shows the set-up 
of  the corresponding model experiment Mini-LIMMCAST \cite{TIMMEL} 
in which a liquid metal 
(in our case GaInSn) is poured from a tundish 
through a submerged entry nozzle 
into the mould. For this configuration 
it is sufficient to apply only one magnetic field by  a 
single coil (see Fig. 4a).
The interaction of the flow with the applied field produces 
induced magnetic fields that we measure at 2x7 
positions at the narrow
faces of the mould in order to reconstruct from them the velocity field.
The mathematics of this inversion relies again on the minimization 
of the mean squared 
deviation of the measured magnetic fields from the fields 
according to the integral equation system (\ref{int1}-\ref{int2}). 
This minimization is done by solving the normal 
equations, whereby we use various auxiliary functionals which serve to 
ensure the divergence-free condition of the velocity,
to enforce its two-dimensionality, and to minimize 
in parallel its mean quadratic value, weighted by some 
regularization parameter.

Figure 4b and 4c show two examples of the reconstructed 
velocity fields resulting from this 
inversion. Note that due to the bubbling of argon into
the liquid metal at the entrance of the submerged entry nozzle, quite different 
flow structures can result, such as the desired double vortex flow (Fig. 4b), but
also the unwanted single vortex flow as shown in Fig. 4c.

\begin{figure}[h!]
\begin{center}
\epsfxsize=13cm{\epsffile{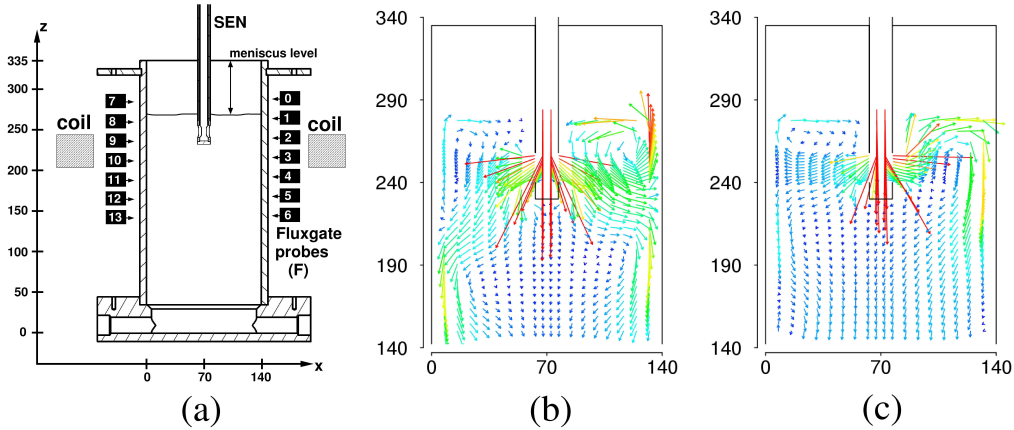}}
\end{center}
\caption{Application of CIFT to a flow problem related to the continuous casting of 
steel. (a) Experimental set-up: A liquid metal 
(GaInSn) is poured through a submerged entry nozzle (SEN) into the mould. 
Added argon gas leads to a highly chaotic flow in the mould. 
A magnetic field of about 2 mT is applied by a coil. The flow induced magnetic
fields are measured with 2x7 fluxgate sensors at the narrow faces of the mold.
(b,c) CIFT reconstructed velocity field in the mould at two different instants, 
with a double vortex 
flow (b), and a single vortex flow (c).
}
\end{figure}

\subsection{Inverse problems at large $Rm$}

When extending the range of inverse problems from low to large
$R_m$, we have to cope with
highly nonlinear inverse problem.  Some progress in its 
treatment has been made
for restricted set-ups, for example by applying the so-called frozen-flux
approximation for the Earth's core which allows to determine (still
with some appropriate regularization) solutions for the  
tangential velocity field components at the core-mantle boundary 
from the measured time dependence of the
radial magnetic field component \cite{ROBERTSSCOTT}. 
Nowadays, such inversions are augmented by 
combinations with three-dimensional geodynamo modeling 
using methods like stochastic inversion \cite{AUBERT} or
data assimilation \cite{JACKSON}.

A complementary sort of restricted models is concerned with the
determination of the radial dependence of the helical turbulence
parameter $\alpha(r)$ from spectral properties, under the (highly
artificial) assumption that $\alpha$ is spherically symmetric. Solving
such inverse spectral dynamo problems by means of an evolutionary
strategy it was possible to obtain such $\alpha(r)$ profiles that lead
to oscillatory dynamo solutions \cite{OSZI}.  This model was later
used to develop simple models of geomagnetic reversals that are just
based on noise triggered jumps and relaxation oscillations in the
vicinity of spectral exceptional points
\cite{PRLREVERSAL,EPSL,GAFD}. On this basis, a first attempt was
further made to infer some of the most essential parameters of the
geodynamo, such as its over-criticality and the turbulent resistivity
from the temporal behavior and the statistical properties of field
reversals \cite{IP3}.
 
We have to admit, however, that apart from those 
special solutions with reduced dimensionality, 
inverse dynamo theory is still
in its infancy.

\section{Conclusions and prospects}

In this paper we have discussed two ways of treating 
induction problems in magnetohydrodynamics, one based on the 
differential equation approach, the other one based on the 
integral equation approach. We have applied both methods 
to various forward and inverse problems at small and large 
magnetic Reynolds numbers.

Much work remains to be done in the field. Concerning the 
forward problem, the inclusion of time-dependent velocity 
fields can lead to surprising effects on the thresholds of 
self-excitation, based on non-normal growth \cite{TILGNER} 
or parametric resonance \cite{GIESECKE2012}.

As for the inverse problems at low $Rm$, externally applied fields
with different frequencies could in principle be employed
for better resolving the velocity field in the depth of the
fluid. Likewise, the use of AC fields might give a chance to combine
CIFT with other tomographic methods, such as the Mutual Inductance
Tomography for the identification of conductivity gradients,
as they are connected with the presence of gas bubbles in liquid metals 
\cite{NATASA}. 

Inverse dynamo problems will remain 
a tough nut in the future, with some interesting prospects 
for inverse spectral methods and data assimilation.

\section*{Acknowledgments}
We thank Deutsche Forschungsgemeinschaft for 
financial support in frame of the Collaborative Research Centre SFB
609 (project A5).

\end{document}